# Stepping closer to pulsed single microwave photon detectors for axions search

A. D'Elia, A. Rettaroli, S. Tocci, D. Babusci, C. Barone, M. Beretta, B. Buonomo, F. Chiarello, N. Chikhi, D. Di Gioacchino, G. Felici, G. Filatrella, M. Fistul, L. G. Foggetta, C. Gatti, E. Il'ichev, C. Ligi, M. Lisitskiy, G. Maccarrone, F. Mattioli, G. Oelsner, S. Pagano, L. Piersanti, B. Ruggiero, G. Torrioli and A. Zagoskin

*Abstract*— **Axions detection requires the ultimate sensitivity down to the single photon limit. In the microwave region this corresponds to energies in the yJ range. This extreme sensitivity has to be combined with an extremely low dark count rate, since the probability of axions conversion into microwave photons is supposed to be very low. To face this complicated task, we followed two promising approaches that both rely on the use of superconducting devices based on the Josephson effect.**

**The first one is to use a single Josephson junction (JJ) as a switching detector (i.e. exploiting the superconducting to normal state transition in presence of microwave photons). We designed a device composed of a coplanar waveguide terminated on a current biased Josephson junction. We tested its efficiency to pulsed (pulse duration 10 ns) microwave signals, since this configuration is closer to an actual axions search experiment. We show how our device is able to reach detection capability of the order of 10 photons with frequency 8 GHz.**

**The second approach is based on an intrinsically quantum device formed by two resonators coupled only via a superconducting qubit network (SQN). This approach relies on quantum nondemolition measurements of the resonator photons. We show that injecting RF power into the resonator, the frequency position of the resonant drop in the transmission coefficient (S21) can be modulated up to 4 MHz.**

**We anticipate that, once optimized, both the devices have the potential to reach single photon sensitivity.**

This work was supported by INFN, project SIMP, and partially supported by EU through FET Open SUPERGALAX project, Grant N.863313, and by University of Salerno through Grants 300391FRB19PAGAN and 300391FRB20BARON. *(Corresponding author: Alessandro D'Elia)*

D. Babusci, M. M Beretta, B. Buonomo, A. D'Elia, D. Di Gioacchino, G. Felici, L. G. Foggetta, C. Gatti, C. Ligi, G. Maccarrone, L. Piersanti, A. Rettaroli, and S. Tocci are with the Laboratori Nazionali di Frascati, INFN, 00044 Roma, Italy (e-mail: danilo.babusci@lnf.infn.it; matteo.beretta@lnf.infn.it; bruno.buonomo@lnf.infn.it; alessandro.delia@lnf.infn.it; daniele.digioacchino@lnf.infn.it; giulietto.felici@lnf.infn.it; luca.foggetta@lnf.infn.it; claudio.gatti@lnf.infn.it; carlo.ligi@lnf.infn.it; giovanni.maccarrone@lnf.infn.it; luca.piersanti@lnf.infn.it; alessio.rettaroli@lnf.infn.it; tocci@lnf.infn.it).
F. Chiarello, F. Mattioli, and G. Torrioli are with the IFN-CNR, via Cineto Romano 42 00156 Rome, Italy, and also with the LNF-INFN, Rome, Italy (e-mail: fabio.chiarello@cnr.it; francesco.mattil@cnr.it; guido.torrioli@cnr.it)
C. Barone and S. Pagano are with the Dipartimento di Fisica "E.R.Caianiello", Università degli Studi di Salerno, 84084 Fisciano, Salerno, Italy, and also with the INFN, Salerno, Italy (e-mail: cbarone@unisa.it; spagano@unisa.it).

G. Filatrella is with the Dipartimento di Scienze e Tecnologie, Università del Sannio, 82100 Benevento, Italy, and also with the INFN, Salerno, Italy (e-mail: filatrella@unisannio.it).
N. Chikhi and M. Lisitskiy are with the CNR-SPIN, Institute for Superconductivity, Innovative Materials and Devices, Pozzuoli, 80078, Italy (email: nassim.chikhi@spin.cnr.it, mikhail.lisitskiy@spin.cnr.it)
B.Ruggero is with the CNR-ISASI, Via Campi Flegrei 34 Ex-Olivetti Building, 80078 Pozzuoli NA (email: berardo.ruggiero@cnr.it)
E. Il'ichev G. Oelsner are with the Leibniz Institute of Photonic Technology, 07745, Jena, Germany (emails: evgeni.ilichev@leibniz-ipht.de, gregor.oelsner@leibniz-ipht.de)
M. Fistul is with Ruhr-University Bochum, 44801, Bochum, Germany (email: Mikhail.Fistoul@ruhr-uni-bochum.de)
A. Zagoskin is with Loughborough University, LE11 3TU, Loughborough, United Kingdom(email: A.Zagoskin@lboro.ac.uk)

## I. Introduction

AXIONS are hypothetical particles that received a burst of interest in the last decades since their discovery would solve two of the great mysteries of physics at the same time: the strong CP (Charge-conjugation Parity) problem in high energy physics and dark matter [1]–[4].
The theory predicts that axions can be converted into microwave photons in presence of a strong magnetic field, [1], [5], renewing the interest for the axions search [6]–[11].
The expected power generated by an axion conversion event is extremely low, therefore a photon detector with single photon sensitivity is necessary.
The branch of circuit quantum electrodynamics based on superconducting devices can in principle provide all necessary elements to realize a single microwave photon detector [12]–[17]. In particular, devices based on the Josephson effect demonstrated to be particularly effective as photon detectors[6], [7], [14], [16], [18]–[23]. Two parallel approaches have been adopted so far: using a Josephson junction (JJ) as a switching detector [7], [18], [19], [21], [23]–[25] and employing a JJ based qubit to perform quantum nondemolition measurements of cavity photons [22].

This work reports the latest experimental results obtained by our group in the attempt to find the optimal solution to fabricate a single microwave photon detector devoted to axions search. We tested both a JJ based switching detector and a Superconducting Qubit Network (SQN).

The main features of these two approaches are discussed in the following to highlight advantages and disadvantages of each method.

We fabricated and tested a Current Biased Josephson Junctions (CBJJ) as switching detector [18]–[20], [26], [27].

The interaction of electromagnetic radiation in the microwave range with a CBJJ causes its switching to the resistive state resulting in an output voltage. However, this kind of devices faces a crossroad that may hinder its photon detection capability. On one hand, to reach the maximal sensitivity, a CBJJ should be biased as close as possible to the point where a single photon triggers the switch to the resistive state, i.e. $I_{bias}$ almost equal to the CBJJ critical current.

On the other hand, $I_{bias}$ should not be too close to the transition point in order to minimize the dark counts. An important aspect to consider for axions search is that all the photons generated need to interact with the detector. In other words, a very efficient transport of the photons to the CBJJ is also required.

In the attempt to overcome all these cross-linked limitations, we designed, produced and characterized a superconductive device based on CBJJ. The device design is very simple and consist of a coplanar waveguide (CPW) terminated with a CBJJ. This device allows us to combine the control over the junction potential barrier height, granted by the CBJJ, with an extremely high photon transmission. Nevertheless the detection of single photons has not been demonstrated yet with our device.

On the other hand, employing qubit based detection scheme that exploits quantum nondemolition measurements has already proven its efficiency as single photon detector [22].

We tested a three ports device composed of a SQN and two resonators.

The device is arranged in a transistor-like geometry where the SQN works as a coupling element between the two perpendicular resonators (tuned to have the same resonance frequency). The idea is that the transmission properties of the device are strongly affected in presence of RF [28] even on the level of single photons. Optimization to enable sensitive detection thus requires studying the resonator-SQN-resonator interaction. The advantage of using a SQN over a single qubit is that of a predicted scaling of the signal to noise ratio as $N$ instead of $\sqrt{N}$, where $N$ is the number of qubit in the network [29]–[31]. However, the device fabrication and design are not as simple as for the switching detector and require a much more complicated scheme from both the fabrication side and the measurement set-up.

In this manuscript we demonstrated that using a CBJJ terminated on a transmission line as a switching detector, it is possible to obtain energy sensitivity down to ~ 5 zJ for radiofrequency (RF) pulses of 10 ns and frequency 8 GHz. In addition, tuning the bias current we show that the CBJJ is activated (i.e. switch to the resistive state) when the number of photons arriving at the junction in a relaxation time ($N_\gamma$) is roughly equal to the number of energy levels in the potential well. With our experimental constraints that we could achieve for different frequencies around 8 GHz is $N_\gamma \approx N_{level} \sim 10$ for RF pulses of duration 10 ns.

For the SQN based device, we also show that injecting RF power into one of the two resonators, it is possible to modulate the frequency position of the transmission of the second resonator up to 4 MHz. This work offers a panoramic view of our latest results obtained in the characterization of these two kind of devices and sets the basis for future performance improvements for both the approaches.

## II. CBJJ switching detector: experimental setup and measurements

### A. Experimental setup

The CPW terminated with the Al/AlOx/Al CBJJ is fabricated at IFN-CNR by means of electron beam lithography and two angles shadow evaporation [32], [33].

The CBJJ area is of 8 µm². The AlOx layer has been obtained oxidizing one of the Al electrodes in pure oxygen at a pressure of 3 mbar for 5 minutes. The nominal critical current is $I_c \sim 3$ µA and capacitance C~ 1 pf. The chip containing the device is mounted on a sample-holder which in turn is fixed on the cold plate of a Leiden CF-CS-110-1000 Fridge (base temperature 8 *mK*). The dc and RF lines are decoupled by a bias-tee.

The dc lines are filtered (LC stage, RCR in "T" configuration resistive twisted wires, powder filters respectively at room temperature, 4 K and 10 mK) with a cut-off below 1 *kHz*. The RF lines are attenuated with two -30 dB elements (at 600 mK and at 10mK stages) and a circulator immediately before the bias-tee. We measure an overall attenuation of -80±1 *dB* as a result of an independent calibration, involving three different microwave lines. The working temperature for this experiment was 15 mK.

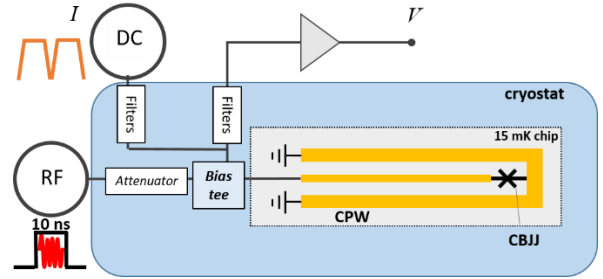

*Figure 1:Schematic representation of the experimental set up.*

### B. Measurements

The arbitrary waveform generated by the Keysight 33500B to current-bias the junction is shown in orange in Fig. 1. The waveform consists in a first ramp until the desired bias value is reached, after which the current is maintained constant. Then, the current is ramped down to some small negative value and the cycle repeats. A TTL signal is generated by the Keysight 33500B at the end of the current ramp-up and triggers the signal generator Rohde&Schwarz SMA100B. The trigger is also sent to the acquisition board NI USB 6366 to start to counting time. After 7 ms from the trigger, the Rohde&Schwarz sends an RF pulse of the desired duration and height. When the junction switches, its voltage raises and from the voltage comparator another TTL is generated and acquired. This is considered an event. The junction has three chances to switch: before the RF pulse, due to the thermal or quantum escape; during the pulse, for the effect of the applied microwaves; after the pulse, again caused by thermal or quantum fluctuations.

The collected lifetime values are defined as the time at which the junction has switched measured starting from the end of the current ramp-up. It might happen that due to a small bias, the escape rate is very low and several ramps are required for an event to occur. To avoid infinite loops in the software, a timeout is set: if after 10 cycles no lifetime is collected, the process is stopped.

As a result, the maximum measurable lifetime is limited to ten times the period of a single current cycles (roughly 140 ms).

## III. CBJJ SWITCHING DETECTOR: RESULTS AND DISCUSSION

### A. Escape Rate measurements without RF

To model our system, we adopt the RCSJ model, where the ideal CBJJ is considered in parallel with a capacitance C and a resistance R [6], [34].

We investigated the switching characteristic of the CBJJ as a function of the bias current in absence of RF. This procedure is useful to investigate the dark counts of the device and to extract useful device parameters like critical current and effective escape temperature.

The switching times are collected in histograms that can be fitted by a simple exponential function:

$$\Delta N(t) = \frac{N_0 \delta t}{\tau} e^{\frac{-t}{\tau}} \qquad (1)$$

Where $\Delta N(t)$ is the number of switching events recorded as a function of the time passed from the bias current application, t, $\frac{N_0 \delta t}{\tau}$ is the number of switching events occurring in the first bin, $\delta t$ is the bin width and $\tau$ is the lifetime. The escape rate $\Gamma = \frac{1}{\tau}$ are calculated as a function of the bias current. The escape rate can be modelled using the Kramer model for thermal activation [35]:

$$\Gamma = a_t \frac{\omega_p}{2\pi} e^{\frac{-\Delta U}{k_b T}} \qquad (2)$$

Where $\frac{\omega_p}{2\pi}$ is the plasma frequency of the CBJJ, $\Delta U$ is the potential well depth, $k_b T$ is the effective temperature multiplied by the Boltzmann constant. $a_t$ is a damping dependent pre-factor and in the case of intermediate-low damping is $a_t = 4/(\sqrt{1 + Q k_b T/1.8 \Delta U} + 1)^2$, where Q is the quality factor of the CBJJ. The expression we used for $\Delta U$, $\omega_p$ and Q are taken

by the seminal work of Martinis, Devoret and Clarke [36]:

$$\Delta U = U_o \left\{ \left[ 1 - \left( \frac{I}{I_c} \right)^2 \right]^{1/2} - \left( \frac{I}{I_c} \right) \cos^{-1} \left( \frac{I}{I_c} \right) \right\} \qquad (3)$$

$$\omega_p = \omega_{p0} \left[ 1 - \left( \frac{I}{I_c} \right)^2 \right]^{1/4} \qquad (4)$$

$$\omega_{p0} = \left( \frac{2\pi I_c}{\phi_0 C} \right)^{\frac{1}{2}} \qquad (5)$$

$$Q = \omega_p RC \qquad (6)$$

Where $U_o = \frac{I_c \phi_0}{2\pi}$, $I = I_{bias}$, $I_c$ is the critical current and $\phi_0$ is the elementary quantum flux. $RC$ is the relaxation time of the CBJJ. The relaxation time is the time necessary to relax to the ground level after the absorption of a photon. In our case R= 50 Ω, the characteristic impedance of the transmission line (TL), and C =1.6 pF, thus $\tau_J \sim 80$ ps.

The experimental escape rate trend as a function of the bias current is reported in Figure 2 along with the best fit obtained using eq. 2.

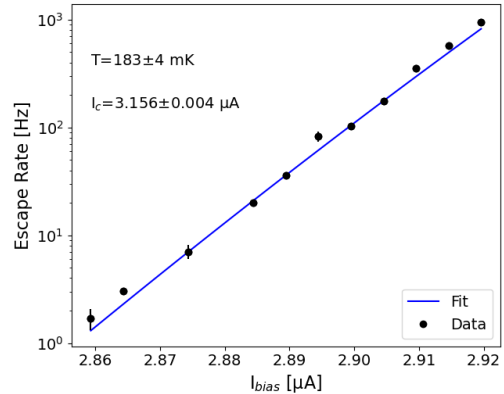

*Figure 2: Escape Rate of the CBJJ as a function of the bias current. Fitting using a Kramer model, the effective temperature of the system and the critical current are estimated to be T=183±4 mK and $I_c$=3.156±0.004 μA.*

From the fitting procedure we obtain an effective temperature of T=183±4 mK and a critical current of $I_c$=3.156±0.004 μA, in good agreement with our previous work [6]. From this analysis we observe that the effective temperature is one order of magnitude larger with respect to the bath temperature (15 mK). This unexpected noise is most likely due to electronic noise not effectively removed by the filters at the 10 mK stage. These filters have been replaced to reduce the effective temperature in future measurements. It is necessary to point out that data reported here have been acquired across multiple days of acquisition. Every day an acquisition like that of figure 2 has been performed to check the CBJJ stability. While the critical current was quite stable around the value of 3.15 μA, the temperature oscillated much more. An average estimation is T=185±15 mK.

The effective temperature of the system is larger than the classical-quantum crossover temperature (~120 mK) [21], [37],

confirming that we are working in the thermal escape regime. Unfortunately, this unexpected level of thermal noise does not allow us to investigate the CBJJ in an ideal regime for single photon detection, where the potential well is 3-4 levels deep. Figure 2 also tells us that the dark count escape rate changes of 1 order of magnitude for each ~0.02 µA of bias current, allowing us to easily reach dark counts rate << 1 Hz by simply reducing the bias. This gives us the opportunity to study the switching efficiency of the CBJJ in two different regimes, high dark counts ($\Gamma$>1 Hz) and low dark counts ($\Gamma$<< 1 Hz). In the high dark counts case, we will investigate bias values that correspond to $\Delta U \sim 10$ levels (each spaced $\hbar\omega_p$) whereas in the low dark counts conditions $\Delta U \sim 20$ levels.

## B. Escape Rate measurements in presence of RF: high dark counts regime

In presence of pulsed RF, the switching times are no longer distributed according to a simple exponential function. When the time difference between the current bias application and the RF arrival on the device is constant, and the RF contribution is concentrated in one single bin, eq. 1 is modified as follow:

$$\Delta N(t) = \frac{N_0\delta t}{\tau}e^{\frac{-t}{\tau}} - N_{RF}e^{\frac{-(t-t_{RF})}{\tau}}\theta(t-t_{RF}) + N_{RF}[\theta(t-t_{RF}) - \theta(t-t_{RF}-\delta t)] \quad (7)$$

Here, $t_{RF}$ (in our case $t_{RF}$ = 7 ms) is the arrival time of the RF, $N_{RF}$ is the number of switching events due to the RF and $\theta$ is the Heaviside function. The first term of eq. 7 is the thermal escape rate before the arrival of the RF. The second term accounts for the reduction of the thermal escape after the RF pulse, while the third set of parenthesis accounts for the extra population in the RF bin. Using this equation, we extract the RF contribution to the escape rate and use it to calculate the switching efficiency of the devices as follows:

$$\varepsilon = \frac{N_{RF}\delta t}{\int_{t_{RF}}^{\infty}\Delta N(t)dt} = \frac{N_{RF}\delta t}{N_0\delta t e^{\frac{-t_{RF}}{\tau}}} \quad (8)$$

$\varepsilon$ is the efficiency, the numerator accounts for all the events in the RF bin only due to the RF presence, the denominator is the number of events that would have been occurred from $t_{RF}$ to infinity, without the RF application.

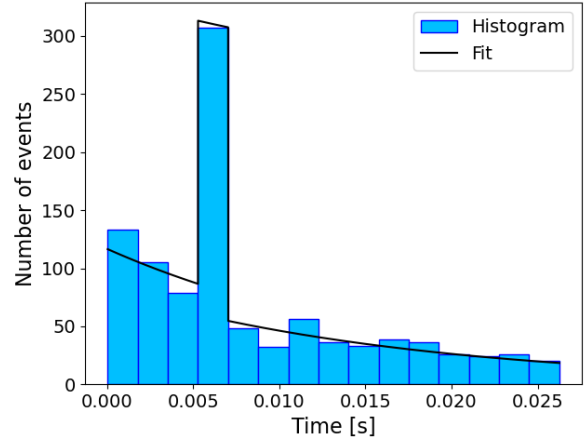

*Figure 3: Histogram of the switching events as a function of time from $I_{bias}$ application, in presence of pulsed RF. The data have been fitted using equation 7 (black line). $I_{bias}$=2.899 µA, RF power $P_{RF}$=-91.58 dbm, RF frequency $\nu_{RF}$=8 GHz, RF pulse width 10 ns.*

We have 4 degrees of freedom to manipulate in order to study the switching efficiency of the device: $I_{bias}$, RF frequency, RF power and RF pulse width (or pulse duration). We studied the switching efficiency behavior changing one of these 4 parameters while keeping the other 3 fixed.

### Variable $I_{bias}$

For these measurements, we keep fixed RF frequency, RF power and RF pulse width while tuning $I_{bias}$. We studied the switching efficiency of the CBJJ in different conditions of RF power and frequency. The RF power is the one that arrives to the CBJJ. For each $I_{bias}$ we acquired the switching histogram as reported in figure 3. Once fitted the data using eq. 7, we used eq. 8 to calculate the efficiency. A typical efficiency curve vs. $I_{bias}$ is reported in figure 4.

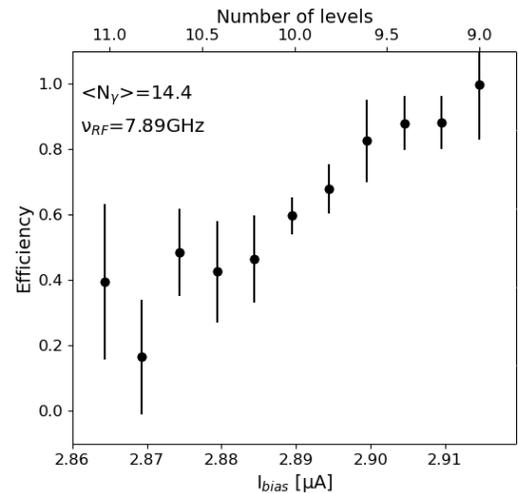

*Figure 4: Switching efficiency vs $I_{bias}$. On the upper x axis are reported the calculated number of levels in the potential well. The RF power, -90.25 dbm, is expressed as $<N_\gamma>$. $<N_\gamma>$ is the average number of photons that arrives at the CBJJ in a relaxation time $\tau_J$~80 ps. RF frequency $\nu_{RF}$=7.89 GHz, RF pulse width = 10 ns. For $I_{bias}$<2.86 µA,*

*equation 8 cannot be used anymore since we are no longer in the high dark counts regime.*

It is convenient to express the RF power as the average number of photons arriving to the CBJJ in a relaxation time instead of dbm. Since after a relaxation time $\tau_J$ the energy stored after the photons absorption is dissipated, we can consider the events happening in two consecutive intervals $\tau_J$ uncorrelated [18]. Knowing the number of photons arriving to the CBJJ in a relaxation time $\tau_J$ is therefore crucial. As well, the potential can also be expressed in number of levels.

$$N_\gamma = \frac{P_{RF}\tau_J}{h\nu_{RF}} \qquad (9)$$

$$N_{level} = \frac{\Delta U(I_{bias})}{\hbar\omega_p\,(I_{bias})} \qquad (10)$$

In equation 9, $N_\gamma$ is the average photon number arriving to the CBJJ in a time interval $\tau_J$, $P_{RF}$ is the RF power, $\tau_J$ (~80 ps) is the RF frequency. In equation 10, $N_{level}$ is the number of levels equi-spaced of $\hbar\omega_p$ in the potential well $\Delta U$, $\omega_p = 2\pi\nu_p$ is the plasma pulsation. In the following, we use $\varepsilon$=0.5 as a reference point. Choosing a value closer to 1 does not affect drastically the obtained result.

From figure 4 it is observable that the CBJJ working point ($\varepsilon$=0.5) corresponds to a number of levels ~10. Considering that we are sending on average ~14 photons in a relaxation time to the junction, the CBJJ is activated when the $N_\gamma \approx N_{levels}$.

This heuristic relation can be intuitively understood through the conservation of energy. If we send to the junction a number of photons $N_\gamma \approx N_{levels}$ with pulsation $\omega_p$, the phase particle will have enough energy to escape the potential well (in first approximation, according to equation 10). This result is reproduced in our simulation [7] in the limit of small $N_{levels}$ (*about* 5) when the CBJJ is galvanically connected to a 50 $\Omega$ TL.

From figure 4. It is also observable that increasing the barrier height, i.e. reducing the bias current, the efficiency decrease substantially. We performed additional measurements changing the RF frequency to test the relation $N_\gamma \approx N_{levels}$. The results are collected in Table 1:

*Table 1: List of $N_\gamma$ and $N_{level}$ at $\varepsilon = 0.5$ for each RF frequency in a 10 ns pulse. The measurements have been performed varying the Rf frequency and the CBJJ $I_{bias}$.*

| $\nu_{RF}$ [GHz] | $N_\gamma$ | $N_{level}@\varepsilon = 0.5$ |
|---|---|---|
| 7.89 | $14.4 \pm^{3.8}_{7.3}$ | 10.3±1.8 |
| 8 | $13.2 \pm^{3.4}_{6.6}$ | 10.8±1.9 |
| 8.04 | $13.4 \pm^{3.5}_{6.7}$ | 9.6±1.7 |
| 8.11 | $14.7 \pm^{3.8}_{7.4}$ | 11.8±2 |
| 8.21 | $14 \pm^{3.6}_{7.1}$ | 8±1.4 |
| 8.51 | $11.5 \pm^{3.0}_{5.8}$ | 8.2±1.4 |

$N_\gamma$ and $N_{level}$ have significant uncertainties, since they depend on the value of the junction capacitance. While the critical current is known with good precision, some ambiguity surrounds the capacitance since different measurements (not shown) of the plasma frequency give different values between 12.1 and 15 GHz. Fixing the critical current to Ic = 3.15 $\mu$A, these values give a capacitance in the range (1-1.6) pF.

The uncertainty on $N_\gamma$ is a combination of both the ambiguity on C and the uncertainty on the calibrated RF power of about 1 dB. The central value of $N_\gamma$ is calculated using C = 1.6 pF.

*Variable RF power $P_{RF}$*

These measurements are characterized by the change of $P_{RF}$ while all the other experimental conditions are fixed. In this case we obtain the efficiency as a function of $N_\gamma$ as reported in figure 5.

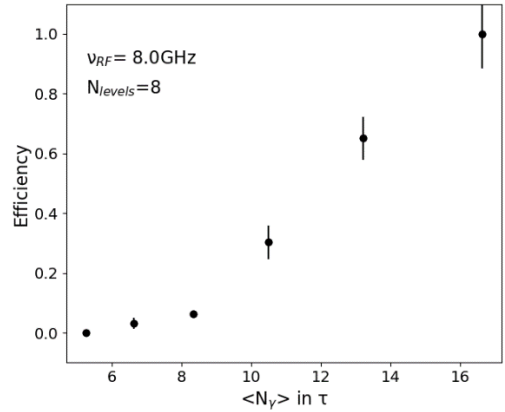

*Figure 5: Switching efficiency as a function of the RF power expressed as $N_\gamma$. $\nu_{RF}$=8 GHz, RF pulse duration = 10 ns, $I_{bias}$=2.899 $\mu$A or equivalently $N_{levels}$=8.*

Also in this case the CBJJ working point corresponds roughly to $N_\gamma = 12$ when the $N_{level} = 8$. This data set also confirms that reducing the number of photons, the CBJJ activation becomes more inefficient. These measurements have been performed also using a frequency of 12 GHz.

The results obtained are reported in Table 2:

*Table 2: List of $N_\gamma$@ $\varepsilon = 0.5$ and $N_{level}$ for each RF frequency in a 10 ns pulse. The measurements have been performed with, $I_{bias}$=2.899 $\mu$A and varying the power sent to the CBJJ.*

| $\nu_{RF}$ [GHz] | $N_\gamma@\varepsilon = 0.5$ | $N_{level}$ |
|---|---|---|
| 8 | $12.0 \pm^{3}_{6}$ | 8±1.4 |
| 12 | $25.0 \pm^{6.5}_{12.6}$ | 8±1.4 |

For an RF frequency of 8 GHz the result is coherent with those in table 1, while for 12 GHz frequency we observe that $N_\gamma/N_{level} \approx 3$, suggesting an influence of the RF frequency over the CBJJ performances. This aspect will be investigated in more detail in future studies.

*Variable RF pulse duration*

For this acquisition we fixed the RF frequency to 8 GHz, the RF power to -92.5 dbm corresponding to $N_\gamma$=8.3, and the $I_{bias}$=2.899 μA, i.e. $N_{level}$ = 8. We changed the pulse duration from 10 ns to 1000 ns, the result is reported in figure 6. We used a Poisson model to describe the data where the switching probability (that is equivalent to the switching efficiency ε treated in the other sections) is written as:

$$\varepsilon = 1 - (1 - \epsilon_J)^{\frac{\Delta T}{\tau_J}} \qquad (11)$$

Where $\varepsilon$ is the switching probability or switching efficiency, $1 - \epsilon_J$ is the probability that the junction does not switch to the resistive state when a single pulse of duration $\tau_J$ is sent to the device ($\epsilon_J$ is therefore the switching efficiency or probability of the device when interacting with a pulse of length $\tau_J$), $\Delta T$ is the pulse width and $\tau_J$ is the relaxation timeof the CBJJ (80 ps). $\Delta T/\tau_J$ represents the number of independent trials within the RF pulse duration.

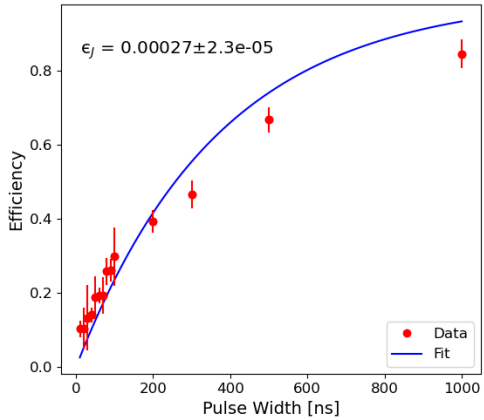

*Figure 6: Switching efficiency ε as a function of the RF pulse duration. The red dots are the data extracted by the switching time distribution, while the blue line is the fit of the data according to equation 11. ε is the switching probability of the CBJJ in a relaxation time $\tau_J$. The poor reduced $\chi^2$=3.38 for the fit, is due to the bias-stability problem discussed in the text.*

Up to $\Delta T = 200$ ns, equation 11 fits the data fairly good. For $\Delta T > 200$ ns the agreement is worse. This is probably due to some experimental drawback that changed the conditions across the measurement. In fact, in this range of pulse width, we recorded a decreasing of the dark counts rate from 65 Hz down to 55 Hz for $\Delta T$=1000 ns. The reason behind this dark counts reduction is not clear at the moment but is consistent with a small drift of the bias current during the measurement. Apart from experimental issues, the trend of the experimental data can be explained by invoking Poisson statistics as already done in the infra-red range and for smaller junctions in the microwave range [18], [38]. The intrinsic switching probability for a pulse of length $\tau_J$ of the CBJJ with these parameters is determined through a fitting procedure to be $\epsilon_J$ =2.7 $10^{-4}$±2.3 $10^{-5}$. With these parameters, the intrinsic switching probability of the CBJJ for a pulse of length $\tau_J$ and to $N_\gamma$=8.3 is determined to be $\epsilon_J$ =2.8 $10^{-4}$±2.3 $10^{-5}$. This corresponds, in figure 5 that shows data acquired in the same experimental conditions, to a switching probability of about 5% in a 10 ns pulse. Such a low power value was chosen to make the switching probability saturate for a pulse length of 1μs. At first order in $\epsilon_J$ equation 11 simplifies as $\epsilon_J \Delta T/\tau_J$. Setting $\Delta T$=10 ns and $\tau_J$=80 ps, $\epsilon_J \Delta T/\tau_J$ = 0.03 in agreement with what observed in figure 5. To our reference point at $\varepsilon = 0.5$ with $N_\gamma$=12 corresponds $\epsilon_J \simeq 1\%$.

These results also set some benchmark values on the energy and power sensitivity of the device. One single photon of 8 GHz has an energy of ~ 5 yJ. Our results show that we are sensitive to ~ 10 photons in a relaxation time, therefore to ~50 yJ in 80 ps. With our experimental set up the shortest pulse we can send is 10 ns long. The average energy arriving to the CBJJ in a 10 ns pulse is therefore $50 \times 10^4$ yJ/$80 \times 10^4$ ps = $50zJ/8 \times 10ns \cong 5zJ/10ns$. We have therefore an energy sensitivity down to ~5 zJ and a power sensitivity down to 5zJ/10ns = 0.5 pW. Considering a quality factor of our device of order 5, the total bandwidth for our device is few GHz, resulting in a noise equivalent power (NEP) of ~ 10 aW/$\sqrt{Hz}$. This result is one order of magnitude larger with respect to the state of the art microbolometers [39]. It is however very promising considering the wide room for optimization of this device.

### B. Escape Rate measurements in presence of RF: low dark counts regime

In this regime we work with $I_{bias}$ in the range 2.55-2.75 μA that corresponds to an expected dark counts escape rate < $10^{-5}$ Hz. In these conditions eq.7 cannot be used. In the switching time distribution, the thermal background is zero and all the counts are concentrated in the RF bin. In this case the efficiency is simply calculated as the ratio between the number of switching events and the total number of switching attempts. In these conditions we measured the switching efficiency as a function of the RF power and the $I_{bias}$, keeping fixed the pulse width to 10 ns and the RF frequency to 8 GHz. The result is showed in fig. 7.

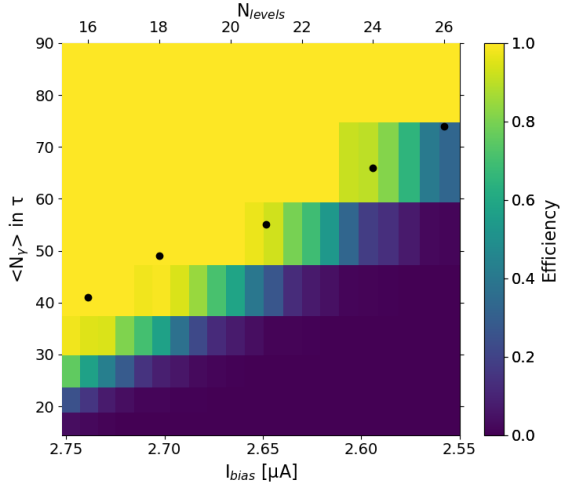

*Figure 7: Switching efficiency as a function of $I_{bias}$ and RF power expressed in photon numbers. The black dots represent the simulation of where the CBJJ is predicted to switch (i.e. efficiency=1). $v_{RF}=8$ GHz, RF pulse width = 10 ns.*

The experimental map is divided into two regions with efficiency 0 (dark blue) and 1 (yellow). The black dots represent the results of the simulations [7]. These, treat the system as purely classical, providing only the values where the efficiencies pass from 0 to 1. A reasonable agreement is observed. Both data and simulation show a large ratio $N_\gamma/N_{level} \approx 2 \div 3$ due to the presence, in the low dark-counts regime, of a deeper potential wells.

The overall performances of the switching detectors tell us that we are able to obtain a sensitivity of ~10 photons/ relaxation time of the junction for frequencies around 8 GHz when the CBJJ potential well is about 10 levels deep. These performances may be improved tuning the RF frequency to match the plasma frequency of the CBJJ. This should considerably improve the device efficiency. In addition, another method potentially very effective is that of artificially increase the quality factor of the device matching the CBJJ and TL impedances by means of a stub tuner. This approach has the purpose of increasing the intrinsic switching time of the junction of a factor proportional to Q, that can be as high as $10^3$. Both these improvements will be tested in the near future.

## IV. SQN THREE PORT DEVICE

A different approach toward single photon detection in the microwave range is that of studying the coupling between SQN and RF resonators.

The main advantage of this approach is that a coupled qubit-resonator device has already been demonstrated to be able to reach single photon sensitivity [22]. An array of interacting qubit, i.e. an SQN, is expected to show a signal to noise ratio scaling that is proportional to the number of element of the SQN[29]–[31], making these kind of devices ideal as single photon detectors. Despite the fact that this approach is more complicated under different aspects with respect to the single JJ based switching detector (e.g. design and fabrication), it is potentially much more efficient in the detection of microwave single photons.

In this section we present preliminary results obtained studying a three ports device (fig. 8) where two geometrically decoupled RF resonators can interact only via the SQN. The idea behind this device is to exploit the readout signal modulation caused by the variation of the Through Transmission of the T resonator (as defined in fig. 8) when an RF signal is applied to the R resonator.

### A. The device

The SQN three ports device was fabricated at Leibniz-IPHT. The SQN design is the so-called T-type device with two resonators at the same resonant frequency coupled by an array of 10 capacitor-shunted flux-qubits. This configuration minimizes unwanted cross-coupling of the resonators. The 25 mm long resonators are fabricated by Nb film on a 200 nm thick silicon substrate. The flux qubits are made of Al Josephson-junctions fabricated by two angle shadow evaporation technique. Every flux qubit of the SQN consists of about 20 μm² loop with three Josephson junctions. Two junctions are designed to have identical size while the third is scaled by a factor α < 1. For qubits of the SQN measured here, the factor α = 0.7. The first resonator, T-Resonator in fig. 8, is delimited by two capacitors of about 5 fF each. The second resonator, R-Resonator in fig. 8, is terminated with one capacitor of about 5 fF on one side and is capacitively coupled to the flux qubits on the other side.

The calculated frequencies of the T-resonator modes are listed in Table 3.

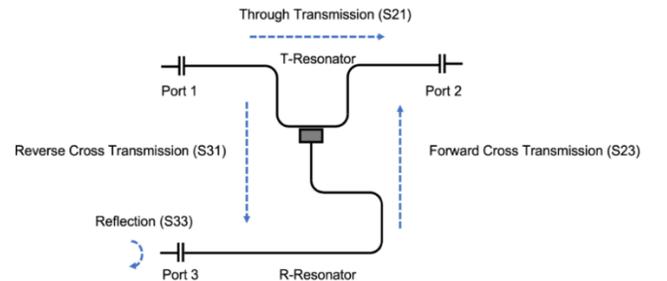

*Figure 8: Schematic picture of the device and of the scattering parameters of the 3 ports device.*

*Table 3: List of the calculated T resonator modes and quality factor*

| Mode | Frequency [GHz] | Loaded Q |
|------|-----------------|----------|
| $\lambda / 2$ | 2.6188 | 56722 |
| $\lambda$ | 5.2376 | 33851 |
| $3\lambda / 2$ | 7.8564 | 23407 |
| $2\lambda$ | 10.4752 | 17788 |

In the following, no flux bias was applied to the SQN.

### B. Scattering parameters characterization.

We first characterized the three-port device by measuring the scattering parameters with the Vector Network Analyzer (VNA). The measurement scheme is shown in fig. 8. We looked for the resonances of the first three modes listed in Table 3. By changing both input port and the output line by the cryogenic switch, we were able to measure both the Trough and Cross transmission coefficients, as well as the reflection on Port 3.

We found strong absorption peaks (resonant drops) in the reflection coefficient on Port 3 (S33), at slightly lower frequencies of those of table 3. We observed peaks at multiples of 2.581 GHz (5.162 GHz and 7.743 GHz).

We argue here that the coupling between the SQN and resonators results in a substantial shifts of resonant frequencies. To prove this, we measured the two-tone spectra as described in the next session.

### C. Two tone spectra of the three-port devices

We considered the third-harmonic absorption-peak of the R-resonator at 7.74 GHz. We set the VNA output-power to -40 dbm, corresponding to about -100 dbm at the device, and measured the through transmission (S21). At the same time, we sent a single tone of frequency 7.743 GHz to the R-resonator with the Rohde&Schwarz SMA100B connected to the Port 3, and varied the output power of the generator from -40 to -20 dbm. By increasing the power sent to Port 3 we clearly see a variation of the resonant-drop frequency in the Through Transmission-spectrum (S21). The data are presented in figure 9a where the frequency shift of the individual S21 spectra as a function of the pump power is clearly observable. The same data is reported in bi-dimensional format in figure 9b. We observe a modulation of the frequency position of the resonant drop of about 4 MHz, proving in principle that the transmitted readout signal can be modulated by an RF signal in the R resonator.

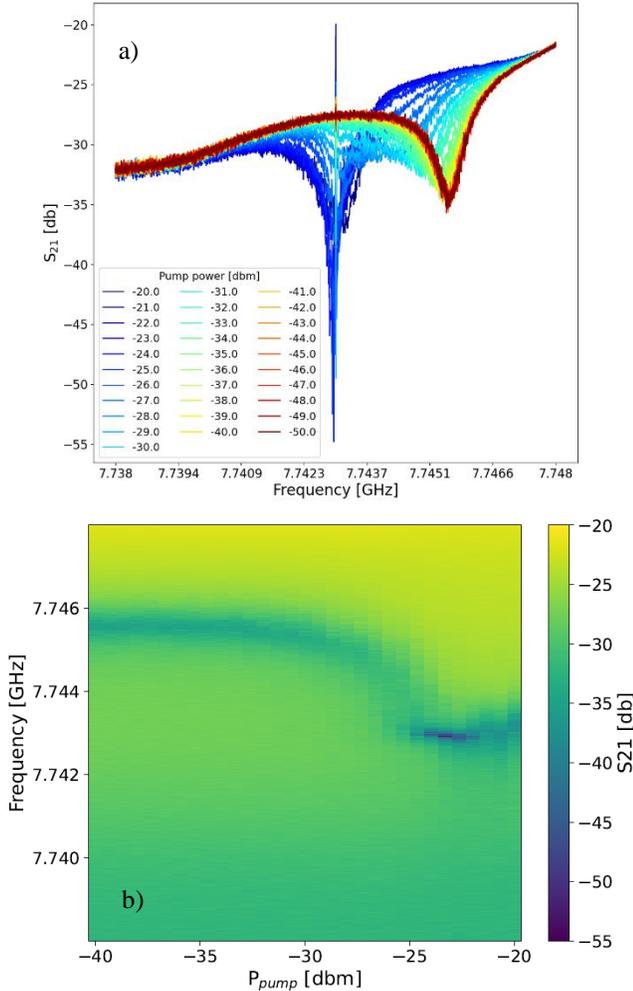

*Figure 9:a) Modulation of the individual through transmission (S21) amplitude acquired with different pump power. RF frequency = 7.743 GHz. b) 2D representation of the graphics in a). S21 frequency modulation as a function of the pump power.*

## V. CONCLUSION

We show the results obtained in the attempt of fabricating a single microwave photon detector using JJ based circuits. We followed two parallel approaches toward single photon detection: the study of a CBJJ based switching detector and the investigation of a SQN based device.

We demonstrate that a switching device composed of a transmission line terminated on a CBJJ, can work as a microwave photon detector for pulsed RF inputs. The capability to tune the potential barrier height by modulating $I_{bias}$, allowed us to demonstrate a sensitivity $N_\gamma \approx N_{level}$. The best performances, constrained by the thermal noise, are reached when $N_{level} \sim 10$ thus obtaining a sensitivity ~ 10 photons for different frequencies around 8 GHz. This result is true for high dark counts rate (>1 Hz), while for low dark count rates (<$10^{-5}$ Hz) where the potential well is deeper we obtain $N_\gamma/N_{level} \approx (2 \div 3)$. When our device is exposed to RF pulses of duration 10 ns we observed an energy sensitivity down to ~5 zJ and a power of 10 aW/Hz$^{1/2}$. This kind of simple design demonstrated itself extremely promising as a photon detector. We believe its performances can be pushed down to single photon sensitivity optimizing it using a stub tuner and tuning the RF frequency to match the CBJJ plasma frequency.

We showed that the SQN three ports device in which two (T and R) low dissipative resonators are electromagnetically coupled through a chain of superconducting qubits, allows to obtain a substantial response to RF signals. The resonant frequency of a T resonator was modulated up to 4 MHz by injecting RF power to the R resonator. We anticipate that once optimized, this device will be a powerful tool in the search for axions.

### ACKNOWLEDGMENTS

We acknowledge funding from the European Union's Horizon 2020 Research and Innovation Program under Grant Agreement No. 863313 (SUPERGALAX).

### REFERENCES

[1]     F. Wilczek, "Problem of Strong PT Invariance in the Presence of Instantons," *Phys. Rev. Lett.*, vol. 40, no. 5, pp. 279–282, Jan. 1978, doi: 10.1103/PhysRevLett.40.279.

[2] R. D. Peccei and H. R. Quinn, "CP Conservation in the Presence of Pseudoparticles," *Phys. Rev. Lett.*, vol. 38, no. 25, pp. 1440–1443, Jun. 1977, doi: 10.1103/PhysRevLett.38.1440.

[3] R. D. Peccei and H. R. Quinn, "Constraints imposed by CPconservation in the presence of pseudoparticles," *Phys. Rev. D*, vol. 16, no. 6, pp. 1791–1797, Sep. 1977, doi: 10.1103/PhysRevD.16.1791.

[4] S. Weinberg, "A New Light Boson?," *Phys. Rev. Lett.*, vol. 40, no. 4, p. 223, Jan. 1978, doi: 10.1103/PhysRevLett.40.223.

[5] P. Sikivie, "Invisible axion search methods," *Rev. Mod. Phys.*, vol. 93, no. 1, p. 015004, Feb. 2021, doi: 10.1103/REVMODPHYS.93.015004/FIGURES/7/MEDIUM.

[6] F. Chiarello *et al.*, "Investigation of Resonant Activation in a Josephson Junction for Axion Search with Microwave Single Photon Detection," *IEEE Trans. Appl. Supercond.*, vol. 32, no. 4, 2022, doi: 10.1109/TASC.2022.3148693.

[7] A. Rettaroli *et al.*, "Josephson Junctions as Single Microwave Photon Counters: Simulation and Characterization," *Instruments*, vol. 5, no. 3, p. 25, Jul. 2021, doi: 10.3390/instruments5030025.

[8] D. Alesini *et al.*, "Search for invisible axion dark matter of mass ma=43 μeV with the QUAX- aγ experiment," *Phys. Rev. D*, vol. 103, no. 10, p. 102004, May 2021, doi: 10.1103/PHYSREVD.103.102004/FIGURES/6/MEDIUM.

[9] N. Crescini *et al.*, "Axion Search with a Quantum-Limited Ferromagnetic Haloscope," *Phys. Rev. Lett.*, vol. 124, no. 17, p. 171801, May 2020, doi: 10.1103/PHYSREVLETT.124.171801/FIGURES/4/MEDIUM.

[10] A. Caldwell *et al.*, "Dielectric Haloscopes: A New Way to Detect Axion Dark Matter," *Phys. Rev. Lett.*, vol. 118, no. 9, p. 091801, Mar. 2017, doi: 10.1103/PHYSREVLETT.118.091801/FIGURES/3/MEDIUM.

[11] B. T. McAllister, G. Flower, E. N. Ivanov, M. Goryachev, J. Bourhill, and M. E. Tobar, "The ORGAN experiment: An axion haloscope above 15 GHz," *Phys. Dark Universe*, vol. 18, pp. 67–72, Dec. 2017, doi: 10.1016/J.DARK.2017.09.010.

[12] A. Opremcak *et al.*, "Measurement of a superconducting qubit with a microwave photon counter," *Science (80-. ).*, vol. 361, no. 6408, pp. 1239–1242, Sep. 2018, doi: 10.1126/SCIENCE.AAT4625/SUPPL_FILE/AAT4625_OPREMCAK_SM.PDF.

[13] D. I. Schuster *et al.*, "Resolving photon number states in a superconducting circuit," *Nature*, vol. 445, no. 7127, pp. 515–518, Feb. 2007, doi: 10.1038/nature05461.

[14] K. Inomata *et al.*, "Single microwave-photon detector using an artificial Λ-type three-level system," *Nat. Commun.*, vol. 7, no. 1, p. 12303, Nov. 2016, doi: 10.1038/ncomms12303.

[15] S. Kono, K. Koshino, Y. Tabuchi, A. Noguchi, and Y. Nakamura, "Quantum non-demolition detection of an itinerant microwave photon," *Nat. Phys. 2018 146*, vol. 14, no. 6, pp. 546–549, Mar. 2018, doi: 10.1038/s41567-018-0066-3.

[16] A. Wallraff *et al.*, "Strong coupling of a single photon to a superconducting qubit using circuit quantum electrodynamics," *Nature*, vol. 431, no. 7005, pp. 162–167, 2004, doi: 10.1038/nature02851.

[17] O. Astafiev, S. Komiyama, T. Kutsuwa, V. Antonov, Y. Kawaguchi, and K. Hirakawa, "Single-photon detector in the microwave range," *Appl. Phys. Lett.*, vol. 80, no. 22, p. 4250, May 2002, doi: 10.1063/1.1482787.

[18] A. L. Pankratov, L. S. Revin, A. V. Gordeeva, A. A. Yablokov, L. S. Kuzmin, and E. Il'ichev, "Towards a microwave single-photon counter for searching axions," *npj Quantum Inf.*, vol. 8, no. 1, p. 61, Dec. 2022, doi: 10.1038/s41534-022-00569-5.

[19] G. Oelsner *et al.*, "Underdamped Josephson junction as a switching current detector," *Appl. Phys. Lett.*, vol. 103, no. 14, p. 142605, Oct. 2013, doi: 10.1063/1.4824308.

[20] L. S. Kuzmin *et al.*, "Single photon counter based on a josephson junction at 14 GHz for searching galactic axions," *IEEE Trans. Appl. Supercond.*, vol. 28, no. 7, Oct. 2018, doi: 10.1109/TASC.2018.2850019.

[21] L. S. Revin *et al.*, "Microwave photon detection by an Al Josephson junction," *Beilstein J. Nanotechnol.*, vol. 11, pp. 960–965, Jun. 2020, doi: 10.3762/bjnano.11.80.

[22] A. V. Dixit *et al.*, "Searching for Dark Matter with a Superconducting Qubit," 2021, doi: 10.1103/PhysRevLett.126.141302.

[23] D. S. Golubev, E. V. Il'Ichev, and L. S. Kuzmin, "Single-Photon Detection with a Josephson Junction Coupled to a Resonator," *Phys. Rev. Appl.*, vol. 16, no. 1, p. 014025, Jul. 2021, doi: 10.1103/PHYSREVAPPLIED.16.014025/FIGURES/3/MEDIUM.

[24] A. Wallraff, T. Duty, A. Lukashenko, and A. V. Ustinov, "Multiphoton Transitions between Energy Levels in a Current-Biased Josephson Tunnel Junction," *Phys. Rev. Lett.*, vol. 90, no. 3, p. 4, Jan. 2003, doi: 10.1103/PHYSREVLETT.90.037003/FIGURES/5/MEDIUM.

[25] V. Fistul, A. Wallraff, and V. Ustinov, "Quantum escape of the phase in a strongly driven Josephson junction," *Phys. Rev. B*, vol. 68, no. 6, p. 060504, Aug. 2003, doi: 10.1103/PhysRevB.68.060504.

[26] A. Poudel, R. McDermott, and M. G. Vavilov, "Quantum efficiency of a microwave photon detector based on a current-biased Josephson junction," *Phys. Rev. B - Condens. Matter Mater. Phys.*, vol. 86, no. 17, p. 174506, Nov. 2012, doi: 10.1103/PHYSREVB.86.174506/FIGURES/5/MEDIUM.

[27] Y. F. Chen *et al.*, "Microwave photon counter based on josephson junctions," *Phys. Rev. Lett.*, vol. 107, no. 21, p. 217401, Nov. 2011, doi: 10.1103/PHYSREVLETT.107.217401/FIGURES/4/MEDIUM.

[28] J. D. Brehm, A. N. Poddubny, A. Stehli, T. Wolz, H. Rotzinger, and A. V. Ustinov, "Waveguide bandgap engineering with an array of superconducting qubits," *npj Quantum Mater. 2021 61*, vol. 6, no. 1, pp. 1–5, Feb. 2021, doi: 10.1038/s41535-021-00310-z.

[29] M. V. Fistul, O. Neyenhuys, A. B. Bocaz, M. Lisitskiy, and I. M. Eremin, "Quantum dynamics of disordered arrays of interacting superconducting qubits: Signatures of quantum collective states," *Phys. Rev. B*, vol. 105, no. 10, p. 104516, Mar. 2022, doi: 10.1103/PHYSREVB.105.104516/FIGURES/8/MEDIUM.

[30] A. M. Zagoskin *et al.*, "Spatially resolved single photon detection with a quantum sensor array," 2013, doi: 10.1038/srep03464.

[31] P. Navez, A. G. Balanov, S. E. Savel'Ev, and A. M. Zagoskin, "Towards the Heisenberg limit in microwave photon detection by a qubit array," *Phys. Rev. B*, vol. 103, no. 6, p. 064503, Feb. 2021, doi: 10.1103/PHYSREVB.103.064503/FIGURES/3/MEDIUM.

[32]    T. A. Fulton and G. J. Dolan, "Observation of single-electron charging effects in small tunnel junctions," *Phys. Rev. Lett.*, vol. 59, no. 1, p. 109, Jul. 1987, doi: 10.1103/PhysRevLett.59.109.

[33]    B. Buonomo *et al.*, "Aluminum single-electron transistors studied at 0.3 K in different transport regimes," *J. Appl. Phys.*, vol. 89, no. 11, p. 6545, Jun. 2001, doi: 10.1063/1.1361236.

[34]    A. Barone and G. Paternò, "Physics and Applications of the Josephson Effect," *Phys. Appl. Josephson Eff.*, Jul. 1982, doi: 10.1002/352760278X.

[35]    H. A. Kramers, "Brownian motion in a field of force and the diffusion model of chemical reactions," *Physica*, vol. 7, no. 4, pp. 284–304, Apr. 1940, doi: 10.1016/S0031-8914(40)90098-2.

[36]    J. M. Martinis, M. H. Devoret, and J. Clarke, "Experimental tests for the quantum behavior of a macroscopic degree of freedom: The phase difference across a Josephson junction," *Phys. Rev. B*, vol. 35, no. 10, p. 4682, Apr. 1987, doi: 10.1103/PhysRevB.35.4682.

[37]    H. F. Yu *et al.*, "Quantum Phase Diffusion in a Small Underdamped Josephson Junction," *Phys. Rev. Lett.*, vol. 107, no. 6, p. 067004, Aug. 2011, doi: 10.1103/PhysRevLett.107.067004.

[38]    G. N. Gol'tsman *et al.*, "Picosecond superconducting single-photon optical detector," *Appl. Phys. Lett.*, vol. 79, no. 6, pp. 705–707, Aug. 2001, doi: 10.1063/1.1388868.

[39]    G. H. Lee *et al.*, "Graphene-based Josephson junction microwave bolometer," *Nature*, vol. 586, no. 7827, pp. 42–46, Oct. 2020, doi: 10.1038/S41586-020-2752-4.